\documentclass{article}
\usepackage[utf8]{inputenc}
\usepackage{amsmath}
\usepackage{amssymb}
\usepackage{cleveref}
\usepackage{accents}
\usepackage{geometry}

\newcommand{\im}{i}
\newcommand{\utilde}[1]{\underaccent{\tilde}{#1}}
\newcommand{\C}{{\mathbb C}}
\newcommand{\R}{{\mathbb R}}

\newcommand{\be}{\begin{eqnarray}}
\newcommand{\ee}{\end{eqnarray}}

\title{Weyl Curvature Evolution System for GR}
\author{Kirill Krasnov and Adam Shaw\\ {}\\
{\it School of Mathematical Sciences, University of Nottingham, NG7 2RD, UK}}

\date{December 2022}

\begin{document}

\maketitle

\begin{abstract}\noindent Starting from the chiral first-order pure connection formulation of General Relativity, we put the field equations of GR in a strikingly simple evolution system form. The two dynamical fields are a complex symmetric tracefree $3\times 3$ matrix $\Psi^{ij}$, which encodes the self-dual part of the Weyl curvature tensor, as well as a spatial ${\rm SO}(3,\C)$ connection $A_a^i$. The right-hand sides of the evolution equations also contain the triad for the spatial metric, and this is constructed non-linearly from the field $\Psi^{ij}$ and the curvature of the spatial connection $A_a^i$. The evolution equations for this pair are first order in both time and spatial derivatives, and so simple that they could have been guessed without a computation. They are also the most natural generalisations of the equations one obtains in the case of the chiral description of Maxwell's theory. We also determine the modifications of the evolution system needed to enforce the "constraint sweeping", so that any possible numerical violation of the constraints present becomes propagating and gets removed from the computational grid. 
\end{abstract}


\section{Introduction}

In 2015 the LIGO gravitational wave observatory made the first direct observation of gravitational waves \cite{LIGOScientific:2016aoc}. By now, there are about a hundred observed events, and the rate of detection is only expected to increase as new gravitational wave observatories become operational. 
On the theoretical front, the first stable evolution of a binary black hole configuration was achieved in 2005, see \cite{Pretorius:2005gq}. Numerical relativity is by now a mature field, see e.g. \cite{Baumgarte:2010ndz}. At the same time, existing numerical schemes take a very long time to run, with up to one month to complete a single simulation. Geometrodynamics in its standard form is difficult and computationally expensive business. 

It is known that General Relativity can be reformulated in many different ways, see e.g. \cite{Krasnov:2020lku}, and it is not impossible that one of such reformulations may hide the potential to be significantly more efficient for numerical simulations. For a review of different formulations from the perspective of numerical relativity see e.g. \cite{Shinkai:2008yb}. The goal of this paper is to explore one of the alternative  formulations of GR in the hope to find a more economic evolution system than the one provided by the standard metric formulation. 

A special feature of four spacetime dimensions is that GR admits a series of chiral formulations. In the physical Lorentzian signature these formulations necessitate the usage of complex-valued objects, but in all signatures they recast the equations of GR in a form that makes many of its hidden properties manifest. In this paper we explore the potential of one of such chiral formulations, namely the chiral first-order pure connection formulation. The main goal of the paper is to recast the field equations of this formalism into an evolution system form, and also suitably gauge-fix it to ensure "constraint sweeping", see below.

The parent of all chiral formulations of GR in four spacetime dimensions is the Plebanski chiral formalism, see \cite{Plebanski:1977zz} for the original paper and e.g.  \cite{Krasnov:2010olp} for a more modern description. This formalism describes GR as the dynamical theory of 3 fields $A^i, \Sigma^i, \Psi^{ij}$, with the dynamics following from the following action:
\be\label{action-Pleb}
S[A,\Sigma,\Psi] = \frac{\im}{16\pi G}\int_M \Sigma^i \wedge F^i - \frac{1}{2}\Psi^{ij} \Sigma^i\wedge \Sigma^j.
\ee
Here $\im$ is the imaginary unit, see below, $G$ is Newton's constant and $F^i = dA^i + (1/2)\epsilon^{ijk} A^j\wedge A^k$ is the curvature of the ${\rm SO}(3)$ connection $A^i$, and $\epsilon^{ijk}$ is the completely anti-symmetric tensor in $\R^3$. The lowercase Latin indices from the middle of the alphabet are "internal" $\R^3$ indices, $i=1,2,3$. This is the action describing the zero cosmological constant GR. The inclusion of a non-zero $\Lambda$ corresponds to adding one more term to the action, but we will not consider non-zero cosmological constant here, as irrelevant for black hole binaries modelling. 

There are 3 field equations arising by minimising the above action. The field $\Psi^{ij}$, which is required to be tracefree, is a Lagrange multiplier field, variation with respect to which produces
\be\label{metricity}
\Sigma^i\wedge \Sigma^j\sim \delta^{ij}.
\ee
This is known as the "metricity constraint" in the literature. It can be shown that this equation implies that the information encoded by $\Sigma^i$ is that of a spacetime metric together with a frame in the space of self-dual 2-forms. The ${\rm SO}(3)$ gauge symmetry manifest in this formalism is precisely the symmetry of orthogonal transformations acting on this frame. The metric that $\Sigma^i$ encodes can be recovered via the formula (\ref{metric-Sigma}). 

The variation with respect to the connection produces
\be
d^A \Sigma^i = d\Sigma^i + \epsilon^{ijk} A^j \wedge \Sigma^k=0.
\ee
It can be shown that when $\Sigma^i$ satisfies (\ref{metricity}) this equation implies that $A^i$ coincides with the self-dual part of the Levi-Civita connection for the metric encoded by $\Sigma^i$. 

Variation with respect to $\Sigma^i$ produces the equation
\be\label{F-Sigma}
F^i = \Psi^{ij} \Sigma^j.
\ee
When the other two equations are satisfied, this is the Einstein equations in disguise. Indeed, when $A^i$ is the self-dual part of the Levi-Civita connection, the curvature $F^i$ is the self-dual part of the Riemann curvature tensor. The above equation says that the self-dual part of the Riemann curvature tensor is self-dual as the 2-form. This is equivalent to the statement that the metric is Ricci-free. The cosmological constant is encoded in this formalism as the trace of the matrix appearing on the right-hand side of (\ref{F-Sigma}), and so it is zero in the version of the formalism we are now discussing. Thus, the above set of equations is equivalent to Einstein equations with zero cosmological constant. It is important to note that when all field equations are satisfied, the field $\Psi^{ij}$ encodes the components of the self-dual part of the Weyl curvature tensor. 

Metrics of different signatures are encoded differently by a triple of 2-forms $\Sigma^i$. Metrics of Euclidean and split signatures are described by real 2-forms $\Sigma^i$, which is related to the fact that the Hodge star on 2-forms squares to plus the identity in these two signatures. Metrics of Lorentzian signature require complex-valued 2-forms $\Sigma^i$. This explains the appearance of the imaginary unit in front of the action (\ref{action-Pleb}). Real Lorentzian metrics arise via (\ref{metric-Sigma}) when $\Sigma^i$ satisfy the following set of reality conditions
\be\label{reality}
 \Sigma^i \wedge \overline{\Sigma^j}=0, \qquad {\rm Re}(\Sigma^i \wedge \Sigma^i) = 0.
 \ee
 With no reality conditions imposed, a triple of complex 2-forms $\Sigma^i$ satisfying $\Sigma^i \wedge \Sigma^j\sim \delta^{ij}$ encodes complex metrics. Reality conditions is one of the most subtle issues of this approach to GR. We will return to discuss them towards the end of the paper. 

The 3+1 decomposition of the above system of equations leads \cite{Jacobson:1988yy} to Ashtekar's Hamiltonian formalism for GR \cite{Ashtekar:1987gu}. This formalism has already been explored from the perspective of numerical relativity, see \cite{Shinkai:1999bm}.

The formulation of GR that is the main subject of this paper arises by "integrating out" the 2-form field $\Sigma^i$ from (\ref{action-Pleb}). This results in the following action
\be\label{action}
S[A,\Psi] = \frac{\im}{32 \pi G} \int \left(\Psi^{-1}\right)^{ij} F^i \wedge F^j.
\ee
This is an action of a diffeomorphism invariant gauge theory, which is built with the help of a Lagrange multiplier field $\Psi$. The field equations of this formulation of GR will be spelled out below. In this formalism it is the curvature 2-forms $F^i$ that encode the metric. When the connection and $\Psi^{ij}$ satisfies its field equations the metric encoded by $F^i$ is Einstein with zero cosmological constant. There are also reality conditions that need to be imposed in order for this formalism to describe real metrics of Lorentzian signature. If desired, the "Lagrange multiplier" field $\Psi^{ij}$ can be "integrated out" from the action (\ref{action}) as well, resulting in a "pure connection" description, see \cite{Capovilla:1989ac} and \cite{Krasnov:2011pp}. As the experience shows, however, this is not the most sensible strategy, and the dynamics resulting from (\ref{action}) is almost always the most tractable one. We will only consider the formulation based on (\ref{action}) in this paper. 

All chiral formalisms have some remarkable properties. One of the almost immediate properties is a clear separation of the conformal mode of the metric from the other nine metric components. In the "parent" Plebanski formalism the perturbation around a general background is encoded in the perturbation $\delta \Sigma^i$, and this can be decomposed into self-dual and anti-self dual parts. It is not hard to see that the conformal mode is encoded by the self-dual part of $\delta \Sigma^i$, while the other nine components of the metric perturbation are stored in the anti-self-dual part of $\delta \Sigma^i$. 

The behaviour of the linearisation of the Plebanski action around an arbitrary (Einstein) background becomes particularly striking in the pure connection formalism, see Section 5.9.5 of \cite{Krasnov:2010olp} and also \cite{Fine:2019tas}. Thus, it can be shown that the linearisation of the Plebanski action around an arbitrary Einstein background, with all linearised fields apart from the connection integrated out, is of an incredibly simple form, schematically $\Psi^{-1} (\partial a)^2$, when a certain gauge is imposed. It is the availability of a certain gauge in the land of connections, together with the incredible simplicity of the arising gauge-fixed action, that suggested to us that the pure connection formalism may also hide potential as an efficient alternative to the standard metric evolution system for GR. The simplicity of the arising evolution equations in this formalism, see below, shows that this hope is at least to some extent realised. 

One of the main results in this paper is that the field equations resulting from (\ref{action}) can be rewritten in the form of evolution equations, resulting in a strikingly simple system. We present these equations already in the Introduction, to hopefully motivate the reader to understand further details.

The new evolution system of GR that follows from (\ref{action}) makes the field $\Psi^{ij}$, which we remind encodes the self-dual part of the Weyl curvature, the main dynamical field. It is for this reason that we have the "Weyl curvature evolution system for GR" as the title of this paper. However, the evolution equation for $\Psi^{ij}$ explicitly contains the spatial ${\rm SO}(3,\C)$ connection $A^i_a$, where $a=1,2,3$ is the spatial 1-form index. For this reason one must also evolve the spatial connection. The evolution system is as follows
\be\label{evol-equations}
D_t \Psi^{ij} - N^a D_a \Psi^{ij} = \im N \epsilon^{klj} \gamma^{ka} D_a \Psi^{il}, \\ \nonumber
D_t A_a^i - \partial_a A_0^i - N^b F^i_{ba} = \im N \Psi^{ij} \gamma^j_a.
\ee
Here $N, N^a$ are the usual lapse and shift (which are real quantities). The objects $D_t, D_a$ are the covariant derivatives with respect to the $A_0^i$ and $A_a^i$ components of the connection respectively. The object $F^i_{ab}= 2\partial_{[a} A^i_{b]} + \epsilon^{ijk} A^j_b A^k_c$ is the curvature of the spatial connection. The object $\gamma^{i a}$ is the (inverse) spatial triad, which is required to be real by the reality conditions that this system must be supplemented with, and which is constructed from the curvature of the spatial connection according to
\be\label{spatial-frame}
\gamma^{ia} := \sqrt{\frac{{\rm det}(\Psi)}{{\rm det}(\tilde{F})}} (\Psi^{-1})^{ij} \tilde{F}^{ja}, \qquad \tilde{F}^{ia}: = \frac{1}{2} \tilde{\epsilon}^{abc} F^i_{bc}, \qquad {\rm det}(\tilde{F}) := \frac{1}{6} \epsilon^{ijk} \utilde{\epsilon}_{abc} \tilde{F}^{ia} \tilde{F}^{jb}\tilde{F}^{kc}.
\ee
Here and in what follows the tilde over a symbol encodes the fact that the object has density weight one, and the tilde under a symbol denotes density weight minus one. There is a single set of constraints that the system (\ref{evol-equations}) must be supplemented with. These are
\be\label{constr}
\gamma^{ia} D_a \Psi^{ij}=0.
\ee

We note that the objects $\Psi^{ij}$ and $A^i_a$ are inherently complex fields, as is in particular manifest from the fact that their evolution equations contain an explicit factor of the imaginary unit on the right-hand side. Nevertheless, the spatial frame $\gamma^{ia}$ is required to be real (in order to produce a real metric of Lorentzian signature), in the sense that $\gamma^{ia}\gamma^{ib}$ is a real symmetric $3\times 3$ tensor. It can be shown that the evolution equations (\ref{evol-equations}) are compatible with these reality conditions in the sense that, if the reality conditions and their time derivatives are imposed at one moment of time, they will remain satisfied at all times. 

From the point of view of evolution equations (\ref{evol-equations}), the objects that are dynamical are the Weyl curvature field $\Psi^{ij}$ and the spatial connection $A^i_a$. All other fields present in (\ref{evol-equations}), namely $N,N^a, A^i_0$ are not dynamical and can be chosen to be what one wishes. Their presence in the evolution system is of course the manifestation of the diffeomorphism and gauge invariance of the theory. 

The evolution equations (\ref{evol-equations}) encode all of the dynamics of GR. Their form is considerably simpler than that of the standard ADM evolution system. In particular, the familiar Hamiltonian and diffeomorphism constraints of GR are absent in this formulation entirely, and are replaced by a single "Gauss" constraint (\ref{constr}). This fact has significant consequences for the problem of numerical constraint violation and gauge-fixing, and will be further dealt with below. Another feature of the system (\ref{evol-equations}) that makes it differ from the ADM evolution system is that it is first order in both the time and spatial derivatives. 

However, the price that one pays for the simplicity is the non-linearity of the spatial frame (\ref{spatial-frame}) as constructed from the curvature of the connection. The other price is the fact that all quantities have become complex-valued and one must impose the reality conditions to recover real GR. This last issue is not a point of concern in "exact" General Relativity, because the reality conditions can be imposed on the initial data, and it is guaranteed that they will be preserved at all times. However, the issue of the reality conditions may become a problem in numerical GR, when it is no longer guaranteed that the reality conditions are preserved by the evolution, and small errors introduced by the discretisation may make the numerical code invalid very quickly. We will come back to the discussion of these issues in the last section. 

The organisation of the remainder of this paper is as follows. We start in Section \ref{sec:Maxwell} by providing an analogy between the new evolution system for GR (\ref{evol-equations}) and the chiral description of Maxwell theory. In the later, it is possible to combine the electric $E^i$ and magnetic $B^i$ fields into a single complex-valued field $\phi^i = B^i + \im E^i$. The Maxwell's equations then take an incredibly simple form of a single evolution equation for the complex field $\phi^i$. What the system (\ref{evol-equations}) achieves for GR is an exact generalisation of this chiral description of spin one to the case of spin two. 

We give more details on the chiral first-order pure connection formalism in Section \ref{sec:chiral-first-order}. We present details of the derivation of the evolution system (\ref{evol-equations}) in Sections \ref{sec:3+1}, \ref{sec:metric} and \ref{sec:evol-system}. We proceed to describe the gauge-fixing that we expect to be most appropriate for the system (\ref{evol-equations}) in Section \ref{sec:gauge-fixing}.

\section{Chiral description of Maxwell's theory}
\label{sec:Maxwell} 

Most of the things we do in the case of gravity have a Maxwell analog. We discuss the much simpler Maxwell case first, as the parallel with it helps to understand the constructions that follow. 

\subsection{Chiral Maxwell}

The 3+1 covariant form of Maxwell's equations follows from the Lagrangian $(F_{\mu\nu})^2$, where $F_{\mu\nu} = 2\partial_{[\mu} A_{\nu]}$ is the field strength of the gauge potential $A_\mu$. In four spacetime dimensions one can add to this Lagrangian an appropriate multiple of $\im \epsilon^{\mu\nu\rho\sigma} F_{\mu\nu} F_{\rho\sigma}$, which is a surface term. This makes the Lagrangian complex, but brings with it some significant simplifications.

The most useful for our purposes version of this trick is its first-order form, in which one introduces an "auxiliary" complex vector field $\phi^i$, which then ends up to be not auxiliary at all, and encode precisely the quantities one is interested in, namely the electric and magnetic fields. 

At the Lagrangian level we proceed as follows. We start by introducing a triple of self-dual 2-forms 
\be
\Sigma^i = \im dt \wedge dx^i + \frac{1}{2} \epsilon^{ijk} dx^j\wedge dx^k.
\ee
We will also need them in the component form, and our differential form conventions are $\Sigma^i = (1/2) \Sigma^i_{\mu\nu} dx^\mu \wedge dx^\nu$. We sometimes omit the symbol of the wedge product, when no confusion can arise and for compactness of the expressions that result. The metric is taken to have the signature mostly plus $\eta_{\mu\nu} = \textrm{diag}(-1,1,1,1)$. The 2-forms $\Sigma^i$ introduced are self-dual, see (\ref{Sigma-duality}) in the orientation $\epsilon^{0123}=+1$. 

Let us consider the following action
\begin{equation}\label{action-M-chiral}
    S[A,\phi] = - \int \Sigma^{i\mu\nu}\phi^i \partial_\mu A_\nu - g^2 (\phi^i)^2.
\end{equation}
Integrating out the "auxiliary" field $\phi^i$ gives 
\be
\phi^i = \frac{1}{2 g^2} \Sigma^{i\mu\nu} \partial_\mu A_\nu,
\ee
substituting which to the action gives 
\be
S[A] = - \frac{1}{4g^2} \int \left( \Sigma^{i\mu\nu}\partial_\mu A_\nu\right)^2.
\ee
We can now use
\be
\Sigma^{i\mu\nu} \Sigma^{i\rho\sigma} = \eta^{\mu\rho} \eta^{\nu\sigma} -  \eta^{\mu\sigma} \eta^{\nu\rho} - \im \epsilon^{\mu\nu\rho\sigma}
\ee
to conclude that
\be
S[A] = - \frac{1}{8g^2} \int (F_{\mu\nu})^2 - \frac{\im}{2} \epsilon^{\mu\nu\rho\sigma} F_{\mu\nu} F_{\rho\sigma}.
\ee
The last term is a total derivative, which shows that the action (\ref{action-M-chiral}) gives an equivalent description of Maxwell theory. 

\subsection{The action in the 1+3 form}

We now perform the 1+3 decomposition of the action (\ref{action-M-chiral}), and obtain the associated field equations in the form of an evolution system. We have
\be\label{action-M-31}
S[A,\phi] = - \int \phi^i ( \im \partial_t A_i -\im  \partial_i A_0 + \epsilon^{ijk} \partial_j A_k) - g^2 (\phi^i)^2.
\ee
Here $\partial_t := \partial/\partial t$ and $\partial_i := \partial/\partial x^i$. The field equations arising from this action are as follows. First, we have the constraint that arises by varying with respect to $A_0$
\be\label{phi-constr}
\partial_i \phi^i =0.
\ee
Second, we have the equation that determines $\phi^i$ and arises by varying the action with respect to this field
\be\label{evol-eqn-a}
2 g^2 \phi^i =  \im \partial_t A_i -\im  \partial_i A_0 + \epsilon^{ijk} \partial_j A_k = B^i + \im E^i,
\ee
where we rewrote the appearing combinations in terms of the electric and magnetic fields.
Finally, the variation of the action with respect to the spatial part of the electromagnetic potential gives
\be\label{evol-eqn-phi}
\im \partial_t \phi^i = \epsilon^{ijk} \partial_j \phi^k.
\ee
The beauty of this last equation is that it is independent of the electromagnetic potentials $A_i, A_0$, and combines the two Maxwell's equations on $B^i, E^i$ into a single complex equation 
\be\label{evol-eqn-M-compl}
\im \dot{\vec{\phi}} =\nabla \times \vec{\phi}.
\ee
on the complex vector $\vec{\phi}= \vec{B}+\im \vec{E}$. We can now interpret the equation (\ref{evol-eqn-phi}) as the evolution equation for $\phi^i$, and (\ref{evol-eqn-a}) as the evolution equation for the electromagnetic potential. In these evolution equation the quantity $A_0$ remains an arbitrary function whose presence there reflects gauge-invariance. The evolution equation for $\phi^i$ must be supplemented by the constraint (\ref{phi-constr}). The interesting feature of this system is that, if one is only interested in the physical fields $E^i, B^i$, it is sufficient to evolve only $\phi^i$, as its evolution equation decouples from that for the potential. 

\subsection{Constraint sweeping}

The above system of equations puts the dynamical equations of Maxwell's theory into the form of an evolution system subject to the single Gauss constraint (\ref{phi-constr}). It is not hard to see from the form of the equation (\ref{evol-eqn-phi}) that if the constraint is satisfied at some moment of time, it will remain satisfied always. Indeed, this follows from the fact that the divergence of the curl is zero, and so the time derivative of $\phi^i$ is divergence-free.

However, any numerical implementation of the above evolution system will unavoidably lead to constraint violation. If this is not dealt with, any numerical code will become unstable. There is a very simple way to deal with this, which is to introduce an additional field whose time evolution is controlled by the would-be constraint. If the constraint is satisfies the additional field remains zero. However, when constraint violation is introduced, the modified system can be arranged in such a way that any constraint violation propagates away with the speed of light. 

Another way to motivate the modification of (\ref{evol-eqn-phi}) is to consider the double time derivative of $\phi^i$. We have
\be
\ddot{\phi}^i = - \im \epsilon^{ijk} \partial_j \dot{\phi}^k = - \epsilon^{ijk} \partial_j \epsilon^{klm} \partial_l \phi^m = - \partial^i \partial_j \phi^j + \Delta \phi^i.
\ee
Only the last term here is the one desired to obtain the $\Box\phi^i=0$ equation. The first term vanishes when the Gauss constraint is satisfied, but in general obscures the desired $\Box$-type structure of the squared evolution equation. 

All these problems are solved in one go if one introduces an new field $\phi$, which has an evolution equation of its own, and which modifies the evolution equation for $\phi^i$. The new system of evolution equations is
\be\label{eqs-M-modified}
\partial_t \phi = \im \partial_i \phi^i, \\ \nonumber
\partial_t \phi^i = - \im (\epsilon^{ijk} \partial_j \phi^k + \partial^i \phi).
\ee
The squaring procedure now gives
\be
\ddot{\phi}^i =- \im \epsilon^{ijk} (\partial_j \dot{\phi}^k + \partial^i \dot{\phi}) = - \partial^i \partial_j \phi^j + \Delta \phi^i + \partial^i \partial_j \phi^j = \Delta \phi^i.
\ee
It is also easy to see that $\ddot{\phi} = \Delta\phi$, and so both $\phi^i,\phi$ propagate with the speed of light. 

The described modification (\ref{eqs-M-modified}) of the Maxwell evolution system are known under the name of "constraint sweeping". The idea is that the constraint can be imposed only initially, and then any numerical violation of the constraint will propagate off the grid, preventing dangerous accumulation of the constraint violation that can render the code unstable. 

\subsection{Lagrangian for gauge-fixing}

The described modification (\ref{eqs-M-modified}) of the evolution system can be obtained from a simple gauge-fixed Lagrangian. Indeed, we supplement the action (\ref{action-M-31}) by the following gauge-fixing term, which is clearly a version of the Lorentz gauge
\begin{equation}\label{gf-M}
    S_{g.f} = \int \phi \eta^{\mu\nu} \partial_\mu A_\nu - g^2 \phi^2 = \int \phi ( - \partial_t A_0 + \partial_i A_i) - g^2 \phi^2.
\end{equation}
It is easy to see that the modifications this addition does to the field equations of Maxwell's theory are precisely as described by (\ref{eqs-M-modified}). Indeed, the field equation that result by varying the Lagrangian with respect to $A_0$ is now precisely the first equation in (\ref{eqs-M-modified}). The field equation that follows by varying with respect to $A_i$ is the second equation in (\ref{eqs-M-modified}). 
Finally, the equation that follows by varying the action with respect to the new field $\phi$ is
\be\label{evol-eqn-A0}
2g^2 \phi = - \partial_t A_0 + \partial_i A_i.
\ee
This can be viewed as an evolution equation for $A_0$, given $\phi, A_i$. Thus, once our dynamical system is gauge-fixed as described there remain no free functions, and $A_0$ evolves together with $A_i$ as dictated by the corresponding equation. 

\subsection{Reality conditions}

It is interesting to discuss the question of the reality conditions that the system of equations (\ref{eqs-M-modified}) must be supplemented with. It is clear that we want the electromagnetic potentials $A_0, A_i$ to be real. It is then clear that the equation (\ref{evol-eqn-A0}) requires the auxiliary field $\phi$ to be real.
Given that $\phi^i = B^i + \im E^i$, and $\partial_i B^i=0$ automatically, we have ${\rm Re}(\partial_i \phi^i)=0$ satisfied automatically. This makes the first of the equations in (\ref{eqs-M-modified}) consistent. The second equation in (\ref{eqs-M-modified}) then splits into its real and imaginary parts $\partial_t \vec{B}=\nabla \times \vec{E}, \partial_t \vec{E} = - \nabla\times \vec{B} - \nabla\phi$. 

So, the reality conditions are  $A_0, A_i$ are real, as well as the induced by the evolution equations reality conditions on $\phi, \phi^i$. The later can be interpreted as the time derivatives of the reality conditions on $A_0, A_i$. These can be stated as $\phi$ is real, as well as a differential condition
\be\label{reality-M}
\nabla \times {\rm Im}(\vec{\phi}) = \partial_t {\rm Re}(\vec{\phi}),
\ee
which is one of the Maxwell's equations, re-stated as a condition linking the real and imaginary parts of the complex vector $\vec{\phi}$. The fact that some of the reality conditions in the Maxwell case can only be stated as a differential relation will have an analog in the gravitational case. 

Below we will see that the action (\ref{action}) is for GR what the action (\ref{action-M-chiral}) is for Maxwell theory. 

\section{Chiral pure connection gravity in the first-order formalism}
\label{sec:chiral-first-order}

Our presentation of the chiral pure connection formalism is very brief, and we only indicate the main interpretational steps. For more details the reader is referred to the book \cite{Krasnov:2020lku}, see in particular Section 6. 

\subsection{Action and field equations}

Chiral Gravity can be described using an $SO(3,\mathbb{C})$ connection $A^i$ and the complex symmetric tracefree matrix of quantities $\Psi^{ij}$.\footnote{ We remind the reader that the Latin indices $i,j,k,...,z$ are the internal ones on which $SO(3,\mathbb{C})$ acts, Latin indices at the start of the alphabet $a,b,c,..,h$ represent 3d spatial coordinates and Greek indices are the spacetime indices.} As already mentioned, the matrix of the quantities $\Psi^{ij}$ is required to be  tracefree  and symmetric \begin{equation}
    tr(\Psi) = \Psi^{kk} = 0, \quad \Psi^{(ij)} = \Psi^{ij}.
\end{equation}
Varying this action with respect to the connection produces
\begin{equation}
   d^A \left( \frac{1}{\Psi^{ij}} F^j \right) = 0 
    \label{eq:CPC_Lagrangian_variation_wrt_A}
\end{equation}
where $d^A$ is the exterior covariant derivative. The notation that have used and will continue to use in what follows is 
\be
\frac{1}{\Psi^{ij}} \equiv (\Psi^{-1})^{ij}.
\ee

Using the Bianchi identity on the curvature form
\begin{equation}
    d^A F^i = 0,
\end{equation}
one can rewrite \cref{eq:CPC_Lagrangian_variation_wrt_A} as
\begin{equation}
    d^A \left( \frac{1}{\Psi^{ij}} \right) F^j = 0.
    \label{eq:CPC_A_equation_of_motion}
\end{equation}
As we will see below, this contains both a constraint and an evolution equation. Varying the action with respect to $\Psi^{ij}$ leads to
\begin{equation}\label{FF-eqn}
    \frac{1}{\Psi^{ik}} F^k \wedge F^l \frac{1}{\Psi^{lj}} \sim \delta^{ij},
\end{equation}
where the Kronecker delta on the right-hand side appears because only the tracefree part of the variation is required to vanish, thus implying that there is only the (arbitrary) trace part on the right-hand side.
Multiplying this by $(\Psi^2)^{ij}$ results in
\begin{equation}
    F^i \wedge F^j \sim \Psi^{ik} \Psi^{kj}.
    \label{eq:CPC_psi_equation_of_motion}
\end{equation}
The proportionality constant is a (different from zero) scalar function that will be fixed below. 

\subsection{Interpretation of the field equations}

The two equations of motion for our system are thus \cref{eq:CPC_A_equation_of_motion} and 
 \cref{eq:CPC_psi_equation_of_motion}. 
 
 The second equation receives the following interpretation. As we will discuss below, a triple of 2-forms $\Sigma^i$ satisfying $\Sigma^i \wedge \Sigma^j\sim \delta^{ij}$ defines the metric (\ref{metric-Sigma}). One obtains real Lorentzian signature metrics when the 2-forms are complex and satisfy the reality conditions (\ref{reality}).
  
 Coming back to the equation \cref{FF-eqn}, we can see that this equation suggest that we define
 \be\label{Sigma-def}
 \Sigma^i := \frac{1}{\Psi^{ij}} F^j.
 \ee
 The object $\Sigma^i$ so defined is constructed algebraically from the fields $\Psi^{ij}$ and the curvature of the connection $A^i$. The equation \cref{eq:CPC_psi_equation_of_motion} then says that this object satisfies $\Sigma^i \wedge \Sigma^j\sim \delta^{ij}$ and thus defines the metric via (\ref{metric-Sigma}).
 
 The interpretation of the first equation \cref{eq:CPC_A_equation_of_motion} is then as follows. It is clear that using the introduced object $\Sigma^i$ it can be rewritten as $d^A \Sigma^i=0$. It can then be shown that, when \cref{eq:CPC_psi_equation_of_motion} holds, this equation implies that $A^i$ coincides with the self-dual part of the Levi-Civita connection for the metric defined by $\Sigma^i$. When this is the case the Einstein equations for the metric defined by $\Sigma^i$ are a simple consequence of (\ref{Sigma-def}). Indeed, when $A^i$ is the self-dual part of the Levi-Civita connection, then $F^i$ is the self-dual part of the Riemann curvature. The definition (\ref{Sigma-def}) can be rewritten as $F^i =\Psi^{ij} \Sigma^j$, and thus implies that the self-dual part of the Riemann curvature is self-dual as the 2-form. This is one of the ways to state the Einstein condition. With $\Psi^{ij}$ being tracefree, the relation $F^i =\Psi^{ij} \Sigma^j$ also implies that the cosmological constant is zero. Thus, all in all, $F^i =\Psi^{ij} \Sigma^j$ is a rewrite of the Ricci-flatness condition, provided both \cref{eq:CPC_A_equation_of_motion} and  \cref{eq:CPC_psi_equation_of_motion} are satisfied. 
 
 The action (\ref{action}) thus gives a chiral, pure connection description of General Relativity with zero cosmological constant. It becomes the description of Lorentzian signature GR when the reality conditions (\ref{reality}) get imposed. 
 
 The field equations \cref{eq:CPC_A_equation_of_motion} and  \cref{eq:CPC_psi_equation_of_motion} can be shown to have the property that the evolution they describe commutes with the reality conditions. Thus, if reality conditions are imposed at a moment of time, they will continue to hold throughout the evolution. For this reason our strategy will be to first understand the space plus time decomposition of the equations \cref{eq:CPC_A_equation_of_motion} and  \cref{eq:CPC_psi_equation_of_motion}, as well as the problems related to the necessary gauge-fixings of this system. Only at the very end we will return to the problem of the reality conditions and comment on the most sensible ways to deal with it. 

\section{Field equations in the 1+3 split form}
\label{sec:3+1}

In this section we start the procedure of casting the field equations into the form of evolution equations. 

\subsection{Notations and conventions}

A 1+3 split of the connection is performed as
\begin{equation}
    A^i = A^i_0 dt + A^i_a dx^a.
\end{equation}
The curvature form can be similarly split
\begin{equation}
    F^i = \frac{1}{2}F^i_{\mu\nu} dx^\mu dx^\nu = F^i_{0a} dt dx^a + \frac{1}{2}F^i_{ab} dx^a dx^b.
\end{equation}
These two components appearing can be written using the connection
\begin{equation}\label{t-partial-A}
    F^i_{0a} = \partial_t A^i_a - D_a A^i_0, \quad F^i_{ab} = \partial_a A^i_b - \partial_b A^i_a + \epsilon^{ijk}A^j_a A^k_b
\end{equation}
where $D_a \chi^i = \partial_a \chi^i + \epsilon^{ijk} A^j_a \chi^k$ is the spatial covariant derivative. A useful description of $F^i_{ab}$ is given by its dual
\begin{equation}\label{tilde-F}
    \tilde{F}^{ia} := \frac{1}{2}\tilde{\epsilon}^{abc} F^i_{bc} \quad \Rightarrow F^i_{ab} = \utilde{\epsilon}_{abc}\tilde{F}^{ic}.
\end{equation}
The determinant of the field $\tilde{F}^{ia}$ is given by
\begin{equation}
    \det{\tilde{F}} := \frac{1}{6} \epsilon^{ijk} \utilde{\epsilon}_{abc} \tilde{F}^{ia} \tilde{F}^{jb} \tilde{F}^{kc}.
\end{equation}
The inverse of $\tilde{F}^{ia}$ can be defined as
\begin{equation}
    \utilde{P}^{i}_a = \frac{1}{2\det{\tilde{F}}}\epsilon^{ijk}\utilde{\epsilon}_{abc} \tilde{F}^{jb} \tilde{F}^{kc},
\end{equation}
and satisfies
\begin{equation}
    \tilde{F}^{ia} \utilde{P}^j_a = \delta^{ij}, \quad \tilde{F}^{ia} \utilde{P}^i_b = \delta^a_b.
\end{equation}
This new variable $\tilde{F}^{ia}$ fully defines the spatial curvature form. Using this parameterisation the equations of motion can be split into the 1+3 formalism.

\subsection{Field equations in the 1+3 form}

The equation \cref{eq:CPC_A_equation_of_motion} is equivalent to
\begin{align}
    \tilde{\epsilon}^{\mu\nu\rho\sigma} D_\nu \left(\frac{1}{\Psi^{ij}} \right) F^j_{\rho\sigma} = 0
\end{align}
To obtain these in 1+3 form, we set the index $\mu=0,a$. The first case results in 
\begin{equation}
    \tilde{\epsilon}^{0abc} D_a \left(\frac{1}{\Psi^{ij}}\right)F^j_{bc} = \tilde{\epsilon}^{abc} D_a \left(\frac{1}{\Psi^{ij}}\right)F^j_{bc}.
\end{equation}
This gives a constraint equation, which is natural to refer to as the Gauss constraint:
\begin{equation}
    C_G^i = D_a \left(  \frac{1}{\Psi^{ij}} \right) \tilde{F}^{ja} \equiv 0.
    \label{eq:Gauss_constraint}
\end{equation}
The case $\mu = a$ gives
\begin{align}
    \tilde{\epsilon}^{a \nu \rho\sigma} D_\nu \left( \frac{1}{\Psi^{ij}} \right) F^j_{\rho\sigma} = \tilde{\epsilon}^{a0bc} D_t \left( \frac{1}{\Psi^{ij}} \right) F^j_{bc} + 2 \tilde{\epsilon}^{ab0c} D_b \left( \frac{1}{\Psi^{ij}} \right) F^j_{0c}.
\end{align}
We use conventions in which $\tilde{\epsilon}^{0abc}=\tilde{\epsilon}^{abc}$, where a single tilde over the object denotes the fact that this object is a density of weight plus one. One thus gets the evolution equation for $\Psi^{ij}$
\begin{equation}
    D_t \left( \frac{1}{\Psi^{ij}} \right) \tilde{F}^{ja} = \tilde{\epsilon}^{abc} D_b \left( \frac{1}{\Psi^{ij}} \right) F^j_{0c}.
    \label{eq:Inv_Psi_Evolution_Equation}
\end{equation}
This can be rewritten in a convenient form by multiplying by $\utilde{P}^k_a$. This gives
\begin{equation}
    D_t \left( \frac{1}{\Psi^{ik}} \right) = \tilde{\epsilon}^{abc} \utilde{P}^k_a D_b \left( \frac{1}{\Psi^{ij}} \right) F^j_{0c},
    \label{eq:Inv_Psi_Equation_Of_Motion}
\end{equation}
which should be interpreted as the evolution equation for (the inverse of) $\Psi^{ij}$. 

We now perform the 1+3 split of the "metricity" equation \cref{eq:CPC_psi_equation_of_motion}. We have
\begin{equation}
    F^i \wedge F^j = \frac{1}{4}\tilde{\epsilon}^{\mu\nu\rho\sigma} F^i_{\mu\nu}F^j_{\rho\sigma} = \frac{1}{2}\tilde{\epsilon}^{0abc} \left[ F^i_{0a}F^j_{bc} + F^i_{bc} F^j_{0a} \right] = 2 F^{(i}_{0a} \tilde{F}^{j)a} \sim \Psi^{ik}\Psi^{kj}.
\end{equation}
Both left and right sides are symmetric in $ij$, the symmetry can be removed on both sides by introducing an antisymmetric term on the right hand side, encoded by $X^i$. The proportionality constant can be dealt with by introducing a scalar field, $\tilde{\mu}$. The equation of motion looks like
\begin{equation}
    F^i_{0a} \tilde{F}^{ja} = \tilde{\mu} \left( \Psi^{ik} \Psi^{kj} + \epsilon^{ijk} X^k \right)
\end{equation}
again multiplying this by $\utilde{P}^j_a$ results in
\begin{equation}\label{F-0a}
    F^i_{0a} = \tilde{\mu}(\Psi^{ik}\Psi^{kj} + \epsilon^{ijk}X^k)\utilde{P}^j_a.
\end{equation}
The two evolution equations are \cref{eq:Inv_Psi_Equation_Of_Motion} and \cref{F-0a}.

\section{Self-dual 2-forms and the metric}
\label{sec:metric}

The goal of this section is to fix our conventions related to the self-duality, and also describe the formula for the spacetime metric as defined by a triple of 2-forms. We then explicitly compute the metric produced by the 2-forms (\ref{Sigma-def}), and thus give interpretation to various objects appearing in the field equations. 

\subsection{Self-dual 2-forms}

Given a frame $e^0, e^i$, the basis in the space of self-dual 2-forms is given by
\be\label{Sigma-frame}
\Sigma^i = \im e^0 \wedge e^i + \frac{1}{2} \epsilon^{ijk} e^j\wedge e^k.
\ee
They are self-dual 
\be\label{Sigma-duality}
\frac{1}{2} \epsilon_{\mu\nu}{}^{\rho\sigma}\Sigma^j_{\rho\sigma} =  \im \Sigma^i_{\mu\nu}.
\ee
in the orientation $\epsilon^{0123}=+1$, and the metric being of signature mostly plus
\be
ds^2 = - (e^0)^2+(e^1)^2+(e^2)^2+(e^3)^2.
\ee
these 2-forms satisfy the following algebra of the quaternions
\be\label{Sigma-algebra}
\Sigma^i_\mu{}^\alpha \Sigma^j_\alpha{}^\nu = - \delta^{ij} \delta_\mu{}^\nu - \epsilon^{ijk} \Sigma^k_\mu{}^\nu,
\ee
where our 2-form conventions are $\Sigma^i= (1/2)\Sigma^i_{\mu\nu} dx^\mu dx^\nu$. 

Further, the 2-forms (\ref{Sigma-frame}) satisfy
\be\label{Sigma-Sigma}
\Sigma^i \wedge \Sigma^j =  2\im \delta^{ij} v_g, \qquad v_g = e^0\wedge e^1\wedge e^2\wedge e^3.
\ee
We take our 4-manifold to be oriented by $v_g$, so that it defines the positive orientation. Using $dx^\mu dx^\nu dx^\rho dx^\sigma = \tilde{\epsilon}^{\mu\nu\rho\sigma} d^4x$ we can then write
\be
v_g = \sqrt{-g} d^4x,
\ee
where $\sqrt{-g}$ is (minus) the determinant of the frame, or the square root of (minus) the determinant of the metric. 

Finally, given a basis $\Sigma^i$ in the space of self-dual 2-forms that satisfies all the relations above, the metric can be recovered via
\be\label{metric-Sigma}
\im g_{\mu\nu}  \sqrt{-g} = \frac{1}{12} \tilde{\epsilon}^{\alpha\beta\gamma\delta} \epsilon^{ijk} \Sigma^i_{\mu\alpha}\wedge \Sigma^j_{\nu\beta} \wedge \Sigma^k_{\gamma\delta}.
\ee
The coefficient on the right-hand side of this expression can be checked by multiplying the whole expression with $g^{\mu\nu}$, and then using (\ref{Sigma-algebra}) and (\ref{Sigma-duality}). 

\subsection{1+3 Metric}

To compute the spacetime metric in the $1+3$ decomposed form, we need to fix our orientation conventions. Given that $v_g= e^0e^1 e^2e^3$ is taken to be the positive orientation, we have $dx^0 dx^1 dx^2 dx^3 = \tilde{\epsilon}^{0123} d^4x = d^4x$. We will also take the positive spatial orientation to be $e^1 e^2 e^3$, which gives $\tilde{\epsilon}^{0abc}= \tilde{\epsilon}^{abc}$.

We now compute the metric produced by the connection. The corresponding orthonormal 2-forms are given by (\ref{Sigma-def}), and we have
\begin{align}
    2i\delta^{ij} v_g =  \Sigma^i \wedge \Sigma^j  = \frac{1}{\Psi^{ik}}\frac{1}{\Psi^{jl}} (F^k_{0a} \tilde{F}^{la} + F^l_{0a} \tilde{F}^{ka}) d^4x= 2\tilde{\mu}\delta^{ij} d^4x,
\end{align}
where we have used (\ref{tilde-F}), as well as (\ref{F-0a}). Now $v_g =  \sqrt{-g}  d^4x$ and we have
\begin{equation}
   \tilde{\mu}=\im \sqrt{-g},
\end{equation}
which identifies the function $\tilde{\mu}$ in \cref{F-0a} with the imaginary unit times the square root of the metric determinant. 

We can now compute all the metric components using (\ref{metric-Sigma}). We have
\be
\im g_{00} \sqrt{-g} =  \frac{1}{6} \tilde{\epsilon}^{abc} \epsilon^{ijk} \Sigma^i_{0a} \Sigma^j_{0b} \Sigma^k_{0c}.
\ee
We can now use (\ref{F-0a}) in the form
\be\label{Sigma-0}
\Sigma^i_{0a}= \frac{1}{\Psi^{ij}} F^j_{0a} = \tilde{\mu} (\Psi^{ij} \utilde{P}^j_a + \im \epsilon^{ijk} (\Psi^{jm} \utilde{P}^m_a) N^k), \qquad N^i := -\im{\rm det}(\Psi) \Psi^{ij} X^j,
\ee
where we have introduced a convenient for the future combination $N^i$, to obtain
\be\label{g00}
g_{00}  =  \tilde{\mu}^2 {\rm det}(\Psi \utilde{P}) (1- N^i N^i),
\ee
where
\be
{\rm det}(\Psi \utilde{P}) = {\rm det}(\Psi) {\rm det}(\utilde{P}), \qquad {\rm det}(\utilde{P})= \frac{1}{6} \tilde{\epsilon}^{abc} \epsilon^{ijk} \utilde{P}^i_a \utilde{P}^j_b \utilde{P}^k_c.
\ee
We also have
\be
\im g_{ab} \sqrt{-g} = - \frac{1}{3{\rm det}(\Psi)} \epsilon^{ijk} F^i_{0(a} F^j_{b)c} \tilde{F}^{kc} + \frac{1}{6{\rm det}(\Psi)} \epsilon^{ijk} \tilde{\epsilon}^{pqr} F^i_{ap} F^j_{bq} F^k_{0r}.
\ee
Using the definition of the $\utilde{P}^i_a$, this can be transformed to
\be
\im g_{ab} \sqrt{-g} = \frac{{\rm det}(\tilde{F})}{{\rm det}(\Psi)}  F^i_{0(a} \utilde{P}^i_{b)}, 
\ee
which gives
\be
g_{ab} =  \frac{{\rm det}(\tilde{F})}{{\rm det}(\Psi)}  \Psi^{ki}\Psi^{kj}\utilde{P}^i_a \utilde{P}^j_b = \frac{1}{{\rm det}(\Psi \utilde{P})}  \Psi^{ki}\Psi^{kj}\utilde{P}^i_a \utilde{P}^j_b = \gamma^i_a \gamma^i_b,
\ee
where we have introduced the triad for the spatial metric given by
\be\label{gamma_a}
\gamma_a^i := \frac{1}{\sqrt{{\rm det}(\Psi \utilde{P})}} \Psi^{ij} \utilde{P}^j_a.
\ee

Finally, we can compute
\be
\im g_{0a} \sqrt{-g} = \frac{1}{6} \epsilon^{ijk}  \Sigma^i_{0b} \Sigma^j_{0a} \frac{1}{\Psi^{kl}} \tilde{F}^{lb} + \frac{1}{6} \epsilon^{ijk} \tilde{\epsilon}^{bcd} \Sigma^i_{0b} \frac{1}{\Psi^{jm}} F^m_{ac} \Sigma^k_{0d}=
 \frac{1}{2} \epsilon^{ijk}  \Sigma^i_{0b} \Sigma^j_{0a} \frac{1}{\Psi^{kl}} \tilde{F}^{lb}.
\ee
Using the expression (\ref{Sigma-0}) and simplifying using the algebra of $\utilde{P}$ and $\tilde{F}$ we get
\be
g_{0a}= -\im \tilde{\mu} N^i \Psi^{ij} \utilde{P}^j_a.
\ee
Comparing with the usual parametrisation of the metric by lapse $N$, shift $N^a$ and the spatial metric $\gamma_{ab}$
\be
g_{\mu\nu} = \left( \begin{array}{cc} - N^2 + N^a N^b \gamma_{ab} & N^b \gamma_{ab} \\ N^b \gamma_{ab} & \gamma_{ab} \end{array}\right),
\ee
we can identify
\be
N^a = \sqrt{-g} \, {\rm det}(\Psi \utilde{P}) N^i \frac{1}{\Psi^{ij}} \tilde{F}^{ja}, \qquad N= \sqrt{-g} \sqrt{{\rm det}(\Psi \utilde{P})}.
\ee
We can now see that the reality conditions saying that the 4-metric is real become the conditions that the objects $\sqrt{-g}, N^i$ and $\Psi^{ij} \utilde{P}^j_a$ are real, modulo possibly an ${\rm SO}(3,\C)$ rotation. Note that separately the quantities $\Psi^{ij}$ and $\utilde{P}^j_a$ are complex-valued, and it is only their product that is supposed to satisfy a reality condition and give rise to the frame of a real spatial metric. 

For future reference we mention that the inverse metric is given by
\be
g^{\mu\nu} = \left( \begin{array}{cc} - N^{-2}  & N^a N^{-2} \\ N^a N^{-2} & \gamma^{ab} - N^{-2} N^a N^b \end{array}\right).
\ee

\subsection{Useful formulas}

Now that the components of the curvature $F^i_{\mu\nu}$ are identified with the metric components, it is useful to rewrite the equation (\ref{Sigma-0}) in terms of the metric. The intermediate useful formulas that we need are
\be\label{useful-formulas}
\sqrt{{\rm det}(\Psi\utilde{P})} = \frac{1}{{\rm det}(\gamma)} = \frac{N}{\sqrt{-g}}, \\ \nonumber
\Psi^{ij} \utilde{P}^j_a = \sqrt{{\rm det}(\Psi\utilde{P})} \gamma^i_a = \frac{N}{\sqrt{-g}} \gamma^i_a, \\
\nonumber
\frac{1}{\Psi^{ij}} \tilde{F}^{ja} = \frac{1}{\sqrt{{\rm det}(\Psi\utilde{P})}} \gamma^{ia} = \frac{\sqrt{-g}} {N}\gamma^{ia}, \\ \nonumber
N^i = \frac{N^a \gamma^i_a}{N}.
\ee
Substituting all this into (\ref{Sigma-0}), and using also $\tilde{\mu}=\im \sqrt{-g}$ we get
\be\label{Sigma-0-useful}
\Sigma^i_{0a}= \frac{1}{\Psi^{ij}} F^j_{0a} = \im N \gamma^i_a - \epsilon_{abc} N^b \gamma^{ic}.
\ee

\section{Rewriting the evolution equations}
\label{sec:evol-system}

The goal of this section is rewrite the evolution equations \cref{eq:Inv_Psi_Equation_Of_Motion} and \cref{F-0a} in the form that makes their interpretation most transparent. In particular, we will make use of the quantities $N, N^a, \gamma^i_a$ that were identified by the metric computation in the previous section. 

\subsection{Rewriting of the $\Psi^{ij}$ equation}

We start by manipulating \cref{eq:Inv_Psi_Equation_Of_Motion} so that it is $\Psi$ that evolves rather than $\Psi^{-1}$. We use
\be
d\left(\frac{1}{\Psi}\right) = - \frac{1}{\Psi}d\Psi \frac{1}{\Psi} ,
\ee
and then multiply the resulting equation by $\Psi$ from the left. This gives
\be
D_t \Psi^{is} \frac{1}{\Psi^{sj}} = \tilde{\epsilon}^{abc} \utilde{P}^j_a D_b \Psi^{is} \frac{1}{\Psi^{sk}} F^k_{0c}.
\ee
We then multiply by another copy of $\Psi$ from the right to get
\be
D_t \Psi^{ij} = \tilde{\epsilon}^{abc} \utilde{P}^s_a \Psi^{sj} D_b \Psi^{il} \frac{1}{\Psi^{lk}} F^k_{0c}.
\ee
The next step is to substitute the expressions (\ref{useful-formulas}) and (\ref{Sigma-0-useful}). This gives
\be
D_t \Psi^{ij} = \frac{N}{\sqrt{-g}} \tilde{\epsilon}^{abc} \gamma^j_a  D_b \Psi^{il} 
 (\im N \gamma^l_c - \epsilon_{cdf} N^d \gamma^{l f}).
 \ee
 We then use 
 \be
\tilde{\epsilon}^{abc} \gamma^i_a \gamma^j_b = {\rm det}(\gamma) \epsilon^{ijk} \gamma^{kc}
 \ee
in the first term, while we expand the product of two epsilons in the second term
\be
\tilde{\epsilon}^{abc} \epsilon_{cdf} = {\rm det}(\gamma) (\delta^a_d \delta^b_f - \delta^a_f\delta^b_d).
\ee
The term where $b$ gets contracted with $f$ vanishes by the Gauss constraint. What remains can be written as 
\begin{equation}
    (D_t - N^a D_a) \Psi^{ij} = i N \epsilon^{klj}\gamma^{ka} D_a \Psi^{il}.
    \label{eq:Psi_evolution_equation}
\end{equation}
It is manifest the right-hand side here is automatically tracefree. It is not hard to see that the anti-symmetric part of the right-hand side is a multiple of the constraint  (\ref{eq:Gauss_constraint}), and so vanishes on the constraint surface. The obtained evolution equation for $\Psi^{ij}$ is very simple, and will be, after appropriate modifications related to gauge-fixing, one of our two main evolution equations. 

It is worth remarking that the obtained evolution equation for $\Psi^{ij}$ is completely analogous to the one we have previous described in the case of Maxwell theory, see (\ref{evol-eqn-phi}). To make the parallel even stronger, we note that in the gauge $A_0^i=0, N^a=0, N=1$, the equation (\ref{eq:Psi_evolution_equation}) can be written as $\partial_t \Psi^{ij} = \im (\nabla \times \Psi)^{ij}$, where $(\nabla \times \Psi)^{ij}:= \epsilon^{klj} \gamma^{ka} D_a \Psi^{il}$ is the generalisation of the usual curl on vector fields to rank two symmetric tracefree tensors. We thus indeed see that the Weyl curvature, which is what encodes the spin two degrees of freedom of the gravitational field, evolves in an exact parallel to how the spin one field $\vec{\phi}=\vec{B} +\im \vec{E}$ evolves in the case of electromagnetism. The main difference in the case of gravity is that the metric has now become dynamical and determined by the potential $A_a^i$. There was no need to evolve the electromagnetic potential in the case of Maxwell, as the equation for $\vec{\phi}$ did not contain the potential. In contrast, in the case of gravity we must evolve the connection together with $\Psi^{ij}$ because the spatial triad that appears on the right-hand side of (\ref{eq:Psi_evolution_equation}) is constructed from the curvature of the connection (and $\Psi^{ij}$). 

\subsection{Rewriting the evolution equation for the connection}

The other evolution equation that we need is that for the connection, see (\ref{F-0a}). It is best to take it in the form (\ref{Sigma-0-useful}). We then multiply the equation by $\Psi$, and note that 
\be
\epsilon_{abc} \Psi^{ij} \gamma^{jc} = \frac{N}{\sqrt{-g}} \epsilon_{abc} \tilde{F}^{ic} = \utilde{\epsilon}_{abc} \tilde{F}^{ic} = F^i_{ab}.
\ee
It is then clear that the evolution equation for the connection can be written as
\be\label{evolution-connection}
D_t A^i_a - \partial_a A^i_0 - N^b F^i_{ba} = \im N \Psi^{ij} \gamma^j_a,
\ee
where we have taken the term involving the shift vector to the left-hand side. This equation clearly interprets the field $\Psi^{ij}$ as the time derivative of the spatial connection $A^i_a$, modulo terms involving $A_0^i$ and $N^a$ and related to gauge. 

We note that the right-hand side of both evolution equations explicitly contains the factor of the imaginary unit, so the evolution system obtained is intrinsically complex. We also note that the Maxwell analog of equation (\ref{evolution-connection}) is (\ref{evol-eqn-a}). Again, there is a direct parallel, apart from the fact that $A_a^i$ has now indices of two different types, and so the right-hand side of (\ref{evolution-connection}) must involve an object that relates them.

\section{Constraint sweeping}
\label{sec:gauge-fixing}

The evolution system we have obtained is a system of two evolution equation, for $\Psi^{ij}$ and for the connection. To understand the evolution equation for $\Psi^{ij}$ in particular, we consider a similar evolution equation in a background of Minkowski metric. 

\subsection{Spin two field in Minkowski spacetime}

As a warm-up, and to contrast the situation in gravity with that in the case of Maxwell field, we assume that the metric is that of Minkowski space, and $\gamma^i_a=\delta^i_a$, so that there is no distinction between the internal and spatial indices. We also assume that shift $N^a=0$ and lapse $N=1$. We get the following evolution equation for $\Psi^{ij}$:
\be\label{psi-evolution-flat}
\partial_t \Psi^{ij} = \im \epsilon^{klj} \partial_k \Psi^{il},
\ee
which should be solved subject to the constraint that $\Psi^{ij}$ is symmetric tracefree and transverse $\partial_j \Psi^{ij}=0$, as this is what the constraint (\ref{eq:Gauss_constraint}) becomes. 

Motivated by the Maxwell example, we look for a modification of this evolution system that introduces a new field $\Phi^i$ whose time derivative is related to the Gauss constraint. We then search for an evolution system that evolves both fields $\Psi^{ij}, \Phi^i$ and is such both fields satisfy the $\Box\Psi^{ij}=\Box \Phi^i=0$ equation. It is not hard to show that there is the unique system that reduces to (\ref{psi-evolution-flat}) when $\Phi^i=0$ and has these properties. It is given by
\be\label{spin-two-flat}
\partial_t \Psi^{ij} = \im \left(\epsilon^{kl(j} \partial_k \Psi^{i)l} + 3 \partial^{(i} \Phi^{j)} - \delta^{ij} \partial_k \Phi^k\right), \\ \nonumber
\partial_t \Phi^i = - \frac{\im}{2} \left( \epsilon^{ijk} \partial^j \Phi^k + \partial^{j} \Psi^{ji}\right).
\ee
We note that the last term in the first equation is needed to make the right-hand side tracefree. Taking another time derivative of the first equation and using both equations one can obtain $\Box\Psi^{ij}=0$. Similarly, taking the time derivative of the second equation and using both equations one obtains $\Box\Phi^i=0$. This shows that the system (\ref{spin-two-flat}) describes a spin two field in Minkowski spacetime, with its $5+3=8$ components. The constraint equations $\partial^j\Psi^{ij}=0$ has been replaced by an evolution equation for 3 new fields $\Phi^i$, so that now all the components of $\Psi^{ij}, \Phi^i$ propagate and satisfy the box equation. This guarantees that any constraint violation that accumulates by some numerical error will propagate away and not accumulate where it occurs. Another desirable feature of "constraint sweeping" is that it allows for any error in the initial conditions (when these are imposed only approximately) to propagate away and not spoil the simulation. This is the desired "constraint sweeping". 

We note that it is essential that the right-hand sides of both equations in (\ref{spin-two-flat}) contain the factor of the imaginary unit. It is these factors that give the right sign for the box equation in Minkowksi spacetime on squaring. This is similar to the situation in the Maxwell case, see (\ref{eqs-M-modified}). 

\subsection{Non-linear version}

There is an obvious non-linear version of the flat evolution system (\ref{spin-two-flat}) that is a modification of (\ref{eq:Psi_evolution_equation}). It is given by
\be\label{evol-eqs-GR-modified}
 ( D_t - N^a D_a) \Psi^{ij} = i N \left[ \epsilon^{kl(j} \gamma^{k a} 
 D_a \Psi^{i)l} + 3 \gamma^{a(i} D_a \Phi^{j)} - \delta^{ij}\gamma^{a k} D_a \Phi^{k} \right],
\\ \nonumber
  ( D_t - N^a D_a)  \Phi^i = -\frac{i N}{2} \left[ \epsilon^{ijk}\gamma^{aj} D_a \Phi^k + \gamma^{aj} D_a \Psi^{ij}  \right].
\ee
We take this to be our main evolution system for $\Psi^{ij}, \Phi^i$, while we will continue to evolve  the spatial connection according to (\ref{evolution-connection}). 

\subsection{Lorentz gauge}

In the Maxwell case the new terms in the evolution equations for $\phi^i$ came from adding the Lorentz gauge term to the Lagrangian, see (\ref{gf-M}). We could follow a similar strategy for gravity. Thus, we could consider adding to the Lagrangian $\Phi^i g^{\mu\nu} \partial_\mu A_\nu^i$. This is a step in the right direction. However, the difficulty that arises is that now the metric is dynamical and is a function of the connection. The variation with respect to the connection will produce terms that are not present in the most natural equations (\ref{evol-eqs-GR-modified}). Thus, it appears to be a better strategy to just postulate the evolution equations (\ref{evol-eqs-GR-modified}) for $\Psi^{ij}, \Phi^i$, and add to them some version of the Lorentz gauge $g^{\mu\nu} \partial_\mu A_\nu^i=0$, or possibly $g^{\mu\nu} \partial_\mu A_\nu^i=\Phi^i$. This will make the $A_0^i$ component of the connection evolve, which is desirable, for one will not need to select this set of functions by hand. However, it is clear that the best form of this evolution equation for $A_0^i$ is likely to depend on how the evolution system we derived behaves under numerics. So, we refrain from trying to fix the form of this equation in the present paper.

\section{Discussion}

In this paper we have rewritten Einstein equations of GR as a system of two evolution equations (\ref{evol-equations}) for the fields $\Psi^{ij}$, which is a complex symmetric tracefree field encoding the self-dual part of the Weyl curvature, and the field $A_a^i$, which is a complex ${\rm SO}(3,\C)$ spatial connection. The equations also involve the usual lapse $N$ and shift $N^a$, which are real, and the temporal component of the connection $A_0^i$, which is complex. These fields do not evolve and can be chosen to be arbitrary functions. The spatial triad $\gamma^i_a$ and its inverse $\gamma^{ia}$ that appear on the right-hand sides of the evolution equations are constructed algebraically (but non-linearly), see (\ref{spatial-frame}) from the field $\Psi^{ij}$ and the curvature $F^i_{ab}$ of the spatial connection. 

The evolution equations (\ref{evol-equations}) are incredibly simple. They directly generalise the evolution equations of the chiral formulation of Maxwell theory, see (\ref{evol-eqn-a}), (\ref{evol-eqn-phi}). In the case of Maxwell, one forms the complex linear combination of the electric and magnetic fields $\vec{\phi}=\vec{B}+\im \vec{E}$. In terms of this field, Maxwell's equations become a single complex evolution equation (\ref{evol-eqn-M-compl}). The evolution equation (\ref{evol-eqn-a}) for the electromagnetic potential can be also read of as the definition of $\vec{\phi}$. The analogy between (\ref{evol-equations}) and the Maxwell case is that the first equation generalises (\ref{evol-eqn-M-compl}), while the second gives the evolution equation for the spatial connection and thus can be read as a definition of $\Psi^{ij}$. The main difference between the gravitational and the Maxwell cases is that in the latter one does not need to evolve the electromagnetic potential as it does not appear in the evolution equation for the $\vec{\phi}$. In the gravity case the field strength of the spatial connection is what determines the metric that appears on the right-hand sides of both equations, and for this reason the spatial connection must be evolved together with $\Psi^{ij}$. At the same time, the structure of the field equations (\ref{evol-equations}) is completely analogous to that of (\ref{evol-eqn-a}), (\ref{evol-eqn-phi}). In fact, one could have guesses the system of equations (\ref{evol-equations}) as the most natural generalisation of the system that arises in the spin one case. 

In both the Maxwell and gravity cases there are constraints. In the Maxwell case this is the Gauss constraint $\partial_i \phi^i=0$, in the gravity case this is similarly the Gauss constraint (\ref{constr}), which is again the most natural generalisation of the Gauss constraint in the case of electromagnetism (and so could have been guessed as well). We find it quite satisfactory that the field equations of GR, in the form of evolution equations, could have been guessed without doing any computations, when one works in the appropriate formalism. This is of course consistent with the fact that GR is the unique theory with "some good properties". The form of the evolution equations (\ref{evol-equations}) tells us that GR is the "most natural" theory describing the dynamics of the spin two field $\Psi^{ij}$ propagating on the metric background determined by an ${\rm SO}(3)$ connection. It is possible that this statement can be strengthened and converted into some form of uniqueness theorem in the connection formalism, but we refrain from developing this line of thought further in this paper. 

In the case of both Maxwell and GR the evolution equations for the field $\vec{\phi}$ and $\Psi^{ij}$ can be modified so as to introduce the "constraint sweeping". In both cases the idea is the same. Instead of trying to impose the Gauss constraint at every step of the evolution, one can introduce a new field ($\phi$ in the Maxwell case and $\Phi^i$ in the gravity case) whose time derivative is a multiple of the Gauss constraint. The right-hand sides of the evolution equations for $\vec{\phi}, \Psi^{ij}$ are then modified by $\phi, \Phi^i$ in such a way that all fields satisfy the box equation. This makes the constraint violation propagating, and any possible constraint error that is created by a numerical scheme will propagate away from the grid. This should lead to numerical stability at least as far as the Gauss constraints are concerned. We postpone detailed studies of the well-posedness of the evolution system ({evol-eqs-GR-modified}), (\ref{evolution-connection}) to future work. 

It is very interesting that in the gravity case there are no analogs of the familiar scalar and vector constraints of the metric formalism of GR. In fact, there is a sense in which these constraints have been solved by the formalism that treats $\Psi^{ij}$ as one of the dynamical variables. This is particularly clear when one compares the formalism of this paper to that of Ashtekar Hamiltonian formulation \cite{Ashtekar:1987gu}. The link from our formalism to that in \cite{Ashtekar:1987gu} is provided by introducing 
\be
\tilde{\gamma}^{ia} = \frac{1}{\Psi^{ij}} \tilde{F}^{ja}
\ee
and treating it as the main dynamical variable rather than $\Psi^{ij}$. One then has to impose the constraints which guarantee that $\Psi^{ij}$ constructed from $\tilde{\gamma}^{ia}$ and $\tilde{F}^{ia}$ is symmetric tracefree. These are precisely the vector and scalar constraints of the Hamiltonian formulation \cite{Ashtekar:1987gu}. Thus, the scalar and vector constraints are solved by the formalism that uses $\Psi^{ij}$ in a single stroke, which is one of its attractive features.

Despite the fact that there are no scalar and vector constraints, diffeomorphisms are still gauge, which is signalled by the presence of the freely specifiable lapse and shift functions in the evolution equations. One can still introduce some gauge-fixing of the diffeomorphisms, and thus constrain the lapse and shift in some way. We leave the problem of determination of the best way of doing this (for numerical purposes) to future work. 

One other issue that needs discussing is the non-linearity of the evolution equations that most strongly manifests itself in the fact that one needs to take the factors of $1/\Psi$ in constructing the spatial frame. An unpleasant feature of this formalism is therefore the fact that whenever the Weyl curvature $\Psi^{ij}$ becomes degenerate, it becomes difficult to interpret the evolution equations. And Weyl curvature can be degenerate in some physically interesting situations. For example, the Weyl curvature matrix $\Psi^{ij}$ of any pp-wave spacetime has a single non-zero eigenvalue, and so is not invertible. At the same time, the matrix $\Psi^{ij}$ for the Schwarzschild solution is invertible. Sufficiently far from the two merging black holes, the spacetime is approximately that of a spherically symmetric Schwarzschild solution together with propagating away gravitational radiation. The Weyl curvature is small, but the matrix $\Psi^{ij}$ should continue to remain invertible. At the same time, the only quantity that is needed for the evolution equations is the spatial triad. When some of the eigenvalues of the Weyl curvature are zero and $\Psi^{ij}$ stops to be invertible the corresponding components of the spatial curvature are also zero. What this signals is just that the corresponding components of the spatial triad take the values they are in the case of Minkowski spacetime. So, it may be possible to give interpretation to the evolution equations (\ref{evol-equations}) even in the case when some of the eigenvalues of $\Psi^{ij}$ are zero and this matrix is not invertible. We will return to this idea elsewhere. Alternatively, in the case of a spherically-symmetric spacetime with gravitational radiation, it should be possible to read off the scalar $\psi_4$ that encodes the gravitational wave signal not too far from the merging black holes so that $\Psi^{ij}$ continues to be invertible. 

We find it quite nice that the evolution system (\ref{evol-equations}) propagates directly the Weyl curvature $\Psi^{ij}$, which is where the observable curvature scalar $\psi_4$ is stored. Thus, the procedure for extraction of the observable gravitational wave template should be much less non-trivial business than it is in the case of the metric formalism. 

Our final comments are on the issue of the reality conditions. We first note that the problem of the reality conditions is already present in the much simpler Maxwell case. Indeed, in that case the reality conditions state that the electromagnetic potential $A_i, A_0$ is real. At the same time, if one interprets the evolution equation (\ref{evol-eqn-a}) as that for the spatial potential $A_i$, there is no guarantee that this will remain real. This will only be the case if at every moment of time the potential determined by the imaginary part of the evolution equation 
\be
\partial_t A_i - \partial_i A_0 = 2 g^2 {\rm Im} (\phi^i)
\ee
is the same as the potential that gets determined by its real part
\be
\epsilon^{ijk} \partial_j A_k = 2g^2 {\rm Re}(\phi^i). 
\ee
This implies the reality condition (\ref{reality-M}) that $\phi^i$ must satisfy. However, it is clear that this reality condition is just contained in the evolution equation (\ref{evol-eqn-M-compl}) for $\phi^i$, and so does not need to be separately imposed. 

Our envisaged strategy of dealing with the reality conditions in the gravity case is similar. In that case the main reality condition states that the spatial frame $\gamma^i_a$ that is constructed from $\Psi^{ij}$ and $F^i_{ab}$ is real, in the sense that the spatial metric $\gamma^i_a \gamma^i_b$ is real. This is a complicated nonlinear set of equations, which is algebraic in $\Psi^{ij}$ but involves first derivatives of the spatial connection. At the same time, it is guaranteed by the link to the Plebanski formalism and then its link to the real metric formulation of GR that this reality condition is compatible with the evolution equations in the sense that one must only preserve this condition (and its time derivative) at a single moment of time, and it will remain satisfied at all times. It is clear that the time derivative of this reality condition is a reality condition on the Weyl curvature $\Psi^{ij}$. It is also clear from general principles that this can be given a form of some second-order in derivatives equation on $\Psi^{ij}$, as the reality condition of $\vec{\phi}$ in the Maxwell case is first order in derivatives. It can then be expected that this equation on $\Psi^{ij}$ that follows from the reality conditions on the connection is contained in the evolution equations, and thus does not need to be separately imposed. All in all, the hope should be that one can evolve the system (\ref{evol-equations}) unconstrained by the reality conditions (and thus only constrain the initial data), as in the Maxwell case, and that the result this produces are physical. Whether this turns out to be case only numerical experimentation can show.

\section*{Acknowledgements} KK was supported by the STFC Consolidated Grant ST/V005596/1.


\begin{thebibliography}{99}

\bibitem{LIGOScientific:2016aoc}
B.~P.~Abbott \textit{et al.} [LIGO Scientific and Virgo],
``Observation of Gravitational Waves from a Binary Black Hole Merger,''
Phys. Rev. Lett. \textbf{116} (2016) no.6, 061102
doi:10.1103/PhysRevLett.116.061102
[arXiv:1602.03837 [gr-qc]].

\bibitem{Pretorius:2005gq}
F.~Pretorius,
``Evolution of binary black hole spacetimes,''
Phys. Rev. Lett. \textbf{95} (2005), 121101
doi:10.1103/PhysRevLett.95.121101
[arXiv:gr-qc/0507014 [gr-qc]].

\bibitem{Baumgarte:2010ndz}
T.~W.~Baumgarte and S.~L.~Shapiro,
``Numerical Relativity: Solving Einstein's Equations on the Computer,''
Cambridge University Press, 2010,
doi:10.1017/CBO9781139193344

\bibitem{Krasnov:2020lku}
K.~Krasnov,
``Formulations of General Relativity,''
Cambridge University Press, 2020,
ISBN 978-1-108-67465-2, 978-1-108-48164-9
doi:10.1017/9781108674652

\bibitem{Shinkai:2008yb}
H.~a.~Shinkai,
``Formulations of the Einstein equations for numerical simulations,''
J. Korean Phys. Soc. \textbf{54} (2009), 2513-2528
doi:10.3938/jkps.54.2513
[arXiv:0805.0068 [gr-qc]].

\bibitem{Plebanski:1977zz}
J.~F.~Plebanski,
``On the separation of Einsteinian substructures,''
J. Math. Phys. \textbf{18} (1977), 2511-2520
doi:10.1063/1.523215

\bibitem{Krasnov:2010olp}
K.~Krasnov,
``Plebanski Formulation of General Relativity: A Practical Introduction,''
Gen. Rel. Grav. \textbf{43} (2011), 1-15
doi:10.1007/s10714-010-1061-x
[arXiv:0904.0423 [gr-qc]].

\bibitem{Jacobson:1988yy}
T.~Jacobson and L.~Smolin,
``Covariant Action for Ashtekar's Form of Canonical Gravity,''
Class. Quant. Grav. \textbf{5} (1988), 583
doi:10.1088/0264-9381/5/4/006

\bibitem{Ashtekar:1987gu}
A.~Ashtekar,
``New Hamiltonian Formulation of General Relativity,''
Phys. Rev. D \textbf{36} (1987), 1587-1602
doi:10.1103/PhysRevD.36.1587

\bibitem{Shinkai:1999bm}
H.~a.~Shinkai and G.~Yoneda,
``Asymptotically constrained and real valued system based on Ashtekar's variables,''
Phys. Rev. D \textbf{60} (1999), 101502
doi:10.1103/PhysRevD.60.101502
[arXiv:gr-qc/9906062 [gr-qc]].


\bibitem{Capovilla:1989ac}
R.~Capovilla, T.~Jacobson and J.~Dell,
``General Relativity Without the Metric,''
Phys. Rev. Lett. \textbf{63} (1989), 2325
doi:10.1103/PhysRevLett.63.2325

\bibitem{Krasnov:2011pp}
K.~Krasnov,
``Pure Connection Action Principle for General Relativity,''
Phys. Rev. Lett. \textbf{106} (2011), 251103
doi:10.1103/PhysRevLett.106.251103
[arXiv:1103.4498 [gr-qc]].


\bibitem{Fine:2019tas}
J.~Fine, K.~Krasnov and M.~Singer,
``Local rigidity of Einstein 4-manifolds satisfying a chiral curvature condition,''
doi:10.1007/s00208-020-02097-z
[arXiv:1910.09790 [math.DG]].

\end{thebibliography}
\end{document}